\documentclass[prb,twocolumn,showpacs,floatfix]{revtex4}
\usepackage{graphicx}
\usepackage{dcolumn}
\usepackage{bm}

\begin{document}

\preprint{}

\title{Terahertz radiation-induced conductivity, Kerr and Faraday angles, and
       spin textures in a two-dimensional electron gas with spin-orbit coupling
       subjected to a high magnetic field and periodic potential}

\author{A.A. Perov, L.V. Solnyshkova, and D.V. Khomitsky\footnote{Corresponding author. E-mail: khomitsky@phys.unn.ru}}

\affiliation{Department of Physics, University of Nizhny Novgorod,
              23 Gagarin Avenue, 603950 Nizhny Novgorod, Russian Federation}

\date{\today}

\begin{abstract}

The terahertz radiation-induced conductivity and dielectric polarization tensors as well as the Faraday and Kerr
rotation angles and the non-equilibrium spin textures are studied for two-dimensional electron gas with strong
spin-orbit coupling subjected to high magnetic field and to tunable periodic potential of a two-dimensional
gated superlattice. It is found that both real and imaginary parts of the frequency-dependent induced conductivity
approach maximum values with sharp and detectable peaks at frequencies corresponding to the
inter-subband transitions between spin-split magnetic subbands. The observed properties of the conductivity
tensor frequency dependence are applied for the description of the Kerr and Faraday rotation angles
which can be used as another experimental tool for describing the electron gas in periodic
structures with significant spin-orbit coupling. The formation of radiation-induced spin textures is predicted
having both in-plane and out-of-plane components with space distribution scale
comparable to the superlattice cell size which can be observed experimentally.

\end{abstract}


\maketitle

\section{Introduction}

In recent years the spin-dependent properties of heterostructures with
Rashba \cite{Rashba} or Dresselhaus \cite{Dresselhaus} spin-orbit coupling (SOC)
have attracted considerable attention in both fundamental and applied areas of
condensed matter physics studying the spin-related phenomena and known
as spintronics. \cite{Awschalom,Zutic,Wu2} The interest in SOC-related effects
in semiconductor physics originates from a promising idea of the spin control
without an external magnetic field variations as it was proposed, for example, by
Datta and Das in their scheme of a spin field-effect transistor.\cite{DattaDas}
Besides the transport measurements and the physics of the spin or charge transfer,
the properties of semiconductor heterostructures with SOC caused by an external electromagnetic
radiation have also been the topic of an extensive research. The attention given to the radiation-induced
properties is natural since, on the one hand, the associated experiments are the usual and reliable tool
for determination of various material parameters in condensed matter physics, and, on another hand,
a proposal and observation of new externally triggered effects in heterostructures is always of interest
for both fundamental problems of condensed matter physics and for the technological
applications. \cite{Awschalom,Zutic,Wu2} These effects are very versatile in their nature and have different
measurable parameters, and here we shall focus mainly on the radiation-induced conductivity tensor,
on the dielectric matrix, and on the associated Faraday and Kerr rotation angles as well as on the excited
non-equilibrium spin textures. The major part of the research in this field considered the metal-based
or magnetic semiconductor structures where the conductivity tensor and Kerr effect have been considered,
\cite{Oppeneer,Wierman,Crooker,Richard,Antonov,Sato,Hamrle,Kato,Reichl,Neudert,Kim,Stroppa,Etz,Winter}
although the magnetooptical properties of the molecular semiconductors\cite{Fronk} and for
the systems in the quantum Hall effect regime have also been studied.\cite{Volkov}
Similar effects in non-magnetic semiconductor structures have also been
explored, including the Faraday \cite{Magarill,Kato} and Kerr \cite{Teppe,Olbrich}
effects as well as their properties in the presence of a significant SOC in two-dimensional electron gas
(2DEG)\cite{Zhang} and the field-induced non-equilibrium spin density.\cite{Bhat,Tarasenko,Pletyukhov,Wu1}

Among the various ways of constructing the heterostructure systems with novel properties of quantum states,
transport, dynamical and spin-related effects the formation of 2DEG with lateral gated superlattice provides
a flexible tool also for the semiconductors with strong SOC. It is well-known that an external electromagnetic
field can generate novel effects including spin current injection\cite{Sherman1,Sherman2}
and spin polarization \cite{Wu1,Wu2}in systems with significant SOC.
This has been also demonstrated in SOC superlattices where the properties
of quantum states, dynamical and transport properties together with the radiation-induced spin textures have
been investigated in 1D superlattices without magnetic field,\cite{K4} and the quantum states and quantization
of Hall conductance together with magnetooptical properties were studied in 2D superlattices at high magnetic
field,\cite{DP,PS} respectively. It was found that the gate control together with the external DC electric
or electromagnetic excitation can provide new possibilities of controllable manipulations of energy spectrum,
charge and spin densities, and the Hall conductivity.
Thus, it is of interest to make a step further and to consider the microwave conductance and dielectric
matrices together with the excited spin textures for 2DEG with SOC subjected to a high magnetic field and to the tunable
periodic potential of a 2D gated superlattice. The properties of energy spectrum and spinor
wavefunctions for such system as well as the DC Hall conductance \cite{DP} and magnetooptical
absorption \cite{PS} have already been studied in detail for both Rashba and Dresselhaus contributions to SOC.
The calculations of the magnetooptical conductivity and dielectric tensors and the discussion of the associated
Faraday and Kerr effects together with the excited spin textures in such structures are the primary goals of
the present paper. We shall focus here on the parameters of the InGaAs/GaAs 2DEG structure with dominating Rashba SOC
and on the frequency range of external electromagnetic radiation which may provide the best characteristics of these
effects for possible experimental observation and technological applications.

The paper is organized as follows: in Sec.II we briefly describe the Hamiltonian and the quantum states
in our system, in Sec.III we calculate and discuss the radiation-induced conductivity and dielectric
matrices which are applied in Sec.IV for the calculation of Kerr and Faraday rotation angles, in Sec.V we focus on
the induced non-equilibrium spin density forming the spin textures, and the concluding remarks are given in Sec.VI.

\section{Hamiltonian and quantum states}

We consider the 2DEG in the $(x,y)$ plane in a InGaAs/GaAs heterostructure with the In content
of around $0.23$, where the electron effective mass and $g$-factor are $m^*=0.05 m_0$ and $g=-4.0$,
respectively, and the SOC is dominated by the Rashba term with a significant amplitude
$\alpha=2.5 \cdot 10^{-9}$ ${\rm eV \cdot cm}$.

The corresponding one-electron Hamiltonian has the following form:\cite{DP}
\begin{eqnarray}
\hat H=\frac{1}{2m^{\ast}}(\hat{\bf p}-e{\bf A}/c)^2\hat E+
\frac{\alpha}{\hbar}[{\bf{z}}\times\hat{\bf{\sigma}}] \cdot
\left(\hat{\bf p} -\frac{e}{c}{\bf A} \right)- \\
-\frac{1}{2}g \mu_B H\hat\sigma_z+V(x,y)\hat E.
\label{Ham}
\end{eqnarray}

Here $\hat p_{x,y}$ are the momentum operator components, $m^{\ast}$ is the electron effective mass,
$\hat\sigma_i$ are the Pauli matrices, $\alpha$ is the strength of Rashba
SO coupling term, $g$ is the Land{\`e} factor, $\mu_B$ is the Bohr magneton, and ${\hat E}$ is the unit matrix.
We use the Landau gauge in which the vector potential of the static magnetic field has the form
${\bf A}=(0,Hx,0)$, and consider a simple form of the periodic superlattice potential
$V(x,y)=V_0 \left(\cos 2\pi x/a + \cos 2\pi y/a \right)$ with amplitude $V_0$ and the superlattice period $a$.
The structure of the Hamiltonian matrix as well as the matrix elements have
been discussed in detail in our previous papers.\cite{DP,PS} It was demonstrated that the
SOC mixes the states of pure Landau levels and results in a doubling of the number of the magnetic
subbands formed under the fixed value $p/q$ of magnetic flux quanta $\Phi/\Phi_0=p/q=|e|Ha^2/2\pi \hbar c$
($p$ and $q$ are prime integers) per unit cell. For a given magnetic flux $p/q$ the spectrum of Hamiltonian
(\ref{Ham}) consists of magnetic subbands on the distance of the order $V_0$ near the corresponding Landau levels,
each being split by the Zeeman and by the SOC terms.

As to the structure of the eigenstate of the Hamiltonian (\ref{Ham}), one can express them as a set
of two-component spinors where each component satisfies the generalized Bloch-Peierls conditions
in the magnetic elementary cell which is the initial superlattice cell multiplied by the factor
of $q$ in $x$ direction if the Landau gauge  ${\bf A}=(0,Hx,0)$ is chosen.
Correspondingly, the magnetic Brillouin zone is determined by inequalities
$-\pi/qa \le k_x \le \pi/qa$, $-\pi/a \le k_y \le \pi/a$, and
a two-component spinor wavefunction has the following form:\cite{DP,PS}

\begin{eqnarray}
\Psi_{\bf k}({\bf r})=\frac{1}{La\sqrt q}\sum_{l=-L/2}^{L/2}\sum_{n=1}^{p}
e^{ik_x(lqa+nqa/p)}e^{2\pi iy(lp+n)/a} \times \\
\times e^{ik_yy}\Bigg[A_n({\bf k})
\Phi_0 \bigg(\frac{x-x_0-lqa-nqa/p}{l_H}\bigg)
\left(
\begin{array}{c}
0 \\
1
\end{array}
\right) + \\
+ B_n({\bf k})\frac{1}{\sqrt{1+D_1^2}}
\left(
\begin{array}{c}
\Phi_0\bigg(\frac{x-x_0-lqa-nqa/p}{l_H}\bigg) \\
-D_1\Phi_1\bigg(\frac{x-x_0-lqa-nqa/p}{l_H}\bigg)
\end{array}
\right)
\Bigg].
\label{psi}
\end{eqnarray}

Here, $D_1=\alpha \sqrt{2}/\bigg(l_H \big(E_0^{+}+\sqrt{{E_0^{+}}^2+
2\alpha^2/l_H^2}\big)\bigg)$, $\Phi_{0,1}[\xi]$ are the simple harmonic oscillator
functions, $l_H=c\hbar/|e|H$ is the magnetic length, $E_0^+=\frac{\hbar\omega_c}{2}+\frac{1}{2}g\mu_B H$
and $\omega_c=|e|H/m^{\ast}c$ is the cyclotron frequency,
$\xi_{\ell n}=(x-x_0-\ell qa-nqa/p)/l_H$, $x_0=c\hbar k_y/|e|H$, and $L$ is the sample size in the $y$ direction
which accounts for the wavefunction norm. The expansion coefficients $A_n({\bf k})$ and $B_n({\bf k})$
in (\ref{psi}) are defined together with the energy eigenvalues during the standard
diagonalization procedure for the Schr\"oedinger equation.

\begin{figure}
  \centering
  \includegraphics[width=85mm]{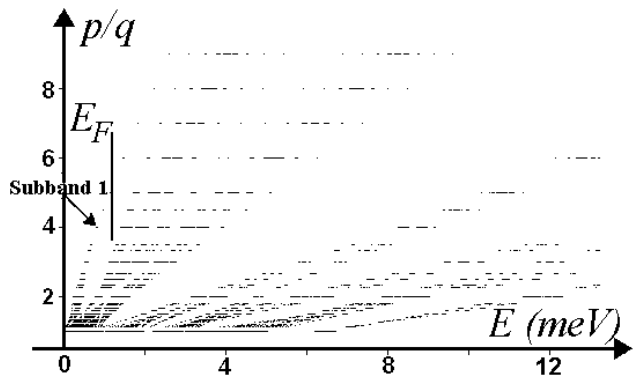}
  \caption{Energy spectrum versus the magnetic flux $p/q$ for two lowest Zeeman and Rashba SOC-split
           Landau levels  the 2DEG in InGaAs/AlGaAs 2D superlattice with Rashba SOC
           $\alpha=2.5 \cdot 10^{-9}$ ${\rm eV \cdot cm}$.
           The electron effective mass $m^{\ast}=0.05$ $m_0$ and the superlattice period
           and amplitude $a=80$ nm and $V_0=1$ meV. Below the effects for subbands at $p/q=4/1$ corresponding
           to the magnetic field $H \approx 2.6$ T are considered with the Fermi level located in the gap above
           the lowest subband 1 marked by arrow.}
  \label{figen}
\end{figure}

An example of the energy spectrum considered
for the following calculations is shown in Fig.\ref{figen} for the magnetic flux
region with $p/q=4/1$ corresponding to $H \approx 2.6$ Tesla and for the superlattice potential amplitude $V_0=1$ meV.
Here the periodic potential amplitude $V_0$ and the SOC energy have
the same order, so the inequality $\Delta E_{SO}\simeq V_0\le \hbar\omega_c$ takes place, and one can use
a two-level approximation where the higher Landau levels do not provide a significant contribution
to the quantum states and the associated effects for two lowest Landau levels.
Below we set $p/q=4/1$ corresponding to the magnetic field $H \approx 2.6$ T in a superlattice with period $a=80$nm,
and the Fermi level is located in the gap above the lowest subband $1$ marked by arrow in Fig.\ref{figen}.
One can see that the photon energy for the inter-subband transitions here is of
the order of $1$ meV corresponding to the frequency $\nu \sim 10^{12}$ $\text{s}^{-1}$
which is the terahertz range being of high interest for the current research in magnetooptical
properties of semiconductor heterostructures.
In a real semiconductor structure with SOC superlattice such position of the Fermi
level corresponds to the concentrations of about $10^{10}\ldots 10^{11}$ ${\rm cm}^{-2}$.
Of course, the manifestation of such tiny and fragile miniband structure as on Fig.\ref{figen}
in measurable experimental phenomena requires the preparation of high-purity samples and low temperatures inside
experimental setups which, as it is known, is still a conventional condition for monitoring the SOC-related
effects in GaAs-based semiconductors where the SOC is relatively weak.

\section{Microwave conductivity and dielectric tensors}

In this Sec. we shall consider the microwave conductivity and dielectric tensors
describing the effects of perturbation caused by the external monochromatic and
linearly polarized electromagnetic field with vector potential ${\bf A}_{\omega}=(A_{\omega},0,0)$
where $A_{\omega}=A_0 \exp (i({\bf k} \cdot {\bf r}-\omega t))$.
Under the dipole approximation the amplitude of the photon wavevector is negligible
compared to the electron quasimomentum which leads to the following form of
the perturbation Hamiltonian ${\hat H}_{\text{int}}$ in the presence of the Rashba SOC:

\begin{equation}
{\hat H}_{\text{int}}=-\frac{i|e|\hbar}{m^{\ast}c}A_0 e^{-i\omega t} \frac{\partial}{\partial x}
-\frac{|e|}{c}\frac{\alpha}{\hbar}A_0 e^{-i\omega t} {\hat \sigma}_y
\label{Hint}
\end{equation}

The main distinguishable property of (\ref{Hint}) compared to the system without SOC
is the presence of second term which is spin-dependent and linear in the Rashba SOC strength.
The SOC also introduces the spin-dependent term into the velocity operator
${\hat {\bf v}}=(i/\hbar)[\hat H, {\bf r}]$ which components ${\hat v}_{x,y}$ in the basis of
the ${\hat \sigma}_z$ eigenstates have the following matrix form:

\begin{equation}
{\hat v}_x=
\left(
\begin{array}{cc}
-i\hbar\nabla_x/m^{\ast} & i\alpha/\hbar \\
-i\alpha/\hbar & -i\hbar\nabla_x/m^{\ast}
\end{array}
\right),
\label{vx}
\end{equation}

\begin{equation}
{\hat v}_y=
\left(
\begin{array}{cc}
-i\hbar\nabla_y/m^{\ast}-\omega_cx & \alpha/\hbar \\
\alpha/\hbar & -i\hbar\nabla_y/m^{\ast}-\omega_cx
\end{array}
\right).
\label{vy}
\end{equation}

With the known eigenfunctions (\ref{psi}) the matrix elements (\ref{vx}), (\ref{vy}) can be expressed
directly via the wavefunction coefficients $A(\bf k)$ and $B(\bf k)$ as

\begin{eqnarray}
v_{{\bf k}\nu\mu}^x=\sum_{n,m=1}^{p}
    \frac{i}{\sqrt{1+D_1^2}}\big(A_n^{\nu\ast}({\bf k})
    B_m^{\mu}({\bf k})- \\
   -B_n^{\nu\ast}({\bf k})A_m^{\mu}({\bf k})\big)
 \cdot\bigg(\frac{D_1\hbar}{m^{\ast}}
    \sqrt{\frac{\pi p}{q}}-\frac{\alpha}{\hbar}\bigg),
\label{vxc}
\end{eqnarray}

\begin{eqnarray}
v_{{\bf k}\nu\mu}^y=\sum_{n,m=1}^{p}
    \frac{1}{\sqrt{1+D_1^2}}\cdot\bigg\{\frac{\alpha}{\hbar}\big(
    A_n^{\nu\ast}({\bf k})B_m^{\mu}({\bf k})+ \\
   +B_n^{\nu\ast}({\bf k})A_m^{\mu}({\bf k})\big)+
\frac{i\hbar D_1}{m^{\ast}l_H\sqrt{2}}\big(A_n^{\nu\ast}({\bf k})B_m^{\mu}
({\bf k})-\hfill \\
 B_n^{\nu\ast}({\bf k})A_m^{\mu}({\bf k})\big)\bigg\}
-\frac{k_y\hbar}{m^{\ast}}\big(A_n^{\nu\ast}({\bf k})
A_m^{\mu}({\bf k}+ \\
+ B_n^{\nu\ast}({\bf k})B_m^{\mu}({\bf k})\big),
\label{vyc}
\end{eqnarray}

which allows to calculate explicitly the components of the conductivity tensor
$\sigma_{ij}(\omega)$ for the frequency range $\nu \sim 0.1 \ldots 1.0$ THz
corresponding to the transitions between the magnetic subband 1 below the Fermi level and the subbands
above the Fermi level shown in Fig.\ref{figen}.
The expression for $\sigma_{ij}(\omega)$ can be derived from the Kubo formula\cite{Kubo}
and is applied for our calculations in the following form: \cite{Oppeneer,Antonov,Stroppa,Kubo,Callaway}

\begin{equation}
\sigma_{ij}(\omega)=\frac {e^2}{8\pi \hbar \omega} \sum\limits_{{\bf k},\mu,\nu}
v_{{\bf k}\mu\nu}^i v_{{\bf k}\nu\mu}^j
f_{{\bf k}\mu}(1-f_{{\bf k}\nu})
\delta (E_{{\bf k}\nu}-E_{{\bf k}\mu}-\hbar \omega),
\label{sigma}
\end{equation}

where (per unit volume) at zero temperature the sum is taken over all states with energy $E_{{\bf k}\mu}$ below and above
the Fermi level, respectively, by applying the Fermi distribution function $f_{{\bf k}\nu}$.
The expression (\ref{sigma}) is written for the absence of scattering,
i.e. for the infinite lifetime of the extended Bloch states.\cite{Stroppa}
It can be verified that this approximation remains qualitatively valid for finite
lifetime also as long as the subband broadening for the spectrum shown in Fig.\ref{figen}
does not exceed the intersubband spacing which can be achieved in clean
heterostructures at low temperatures where the manifestation of the magnetic subband
fine structure is visible.\cite{DP}
It should be mentioned also that the longitudinal part $\sigma_{xx}$ of (\ref{sigma}) is real
while the off-diagonal part $\sigma_{xy}$ has both real and imaginary parts which is a usual
property of the radiation-induced conductivity.

The knowledge of the conductivity tensor (\ref{sigma}) allows us to calculate directly the components
of the frequency-dependent dielectric tensor $\varepsilon_{ij}(\omega)$ via a conventional relation

\begin{equation}
\varepsilon_{ij}(\omega)=\delta_{ij}+\frac{4\pi i}{\omega}\sigma_{ij}({\omega})
\label{diel}.
\end{equation}

\begin{figure}
  \centering
  \includegraphics[width=85mm]{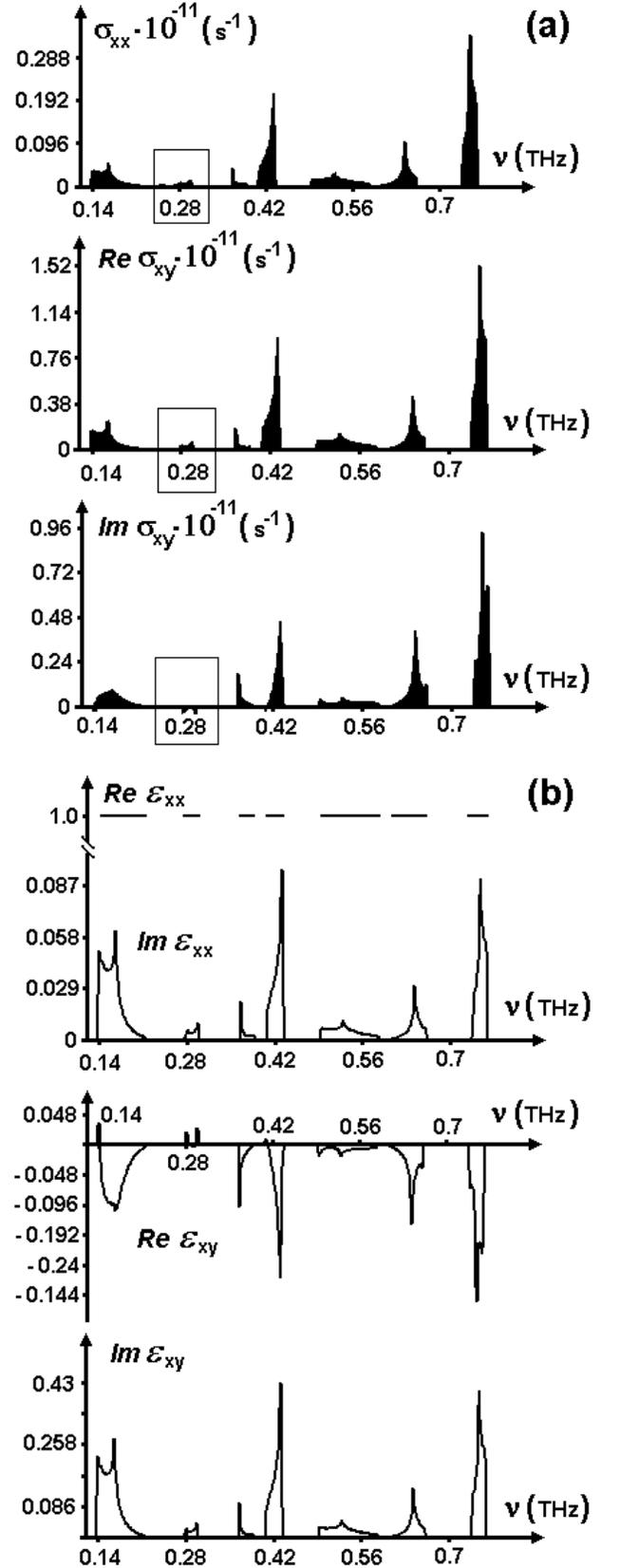}
  \caption{Frequency dependence of (a) $\sigma_{xx}$, real and imaginary parts of $\sigma_{xy}$
           and (b) the same for $\varepsilon_{xx}$, $\varepsilon_{xy}$ for the incident radiation linearly
           polarized along $x$ axis and propagating perpendicular to the 2DEG.
           The components of both the conductivity and the dielectric tensors have their maxima
           at frequencies corresponding to the distance between centers of the magnetic subbands
           shown in Fig.\ref{figen}. The detailed structure near $\nu=0.28$ THz marked
           by rectangle is discussed below.}
  \label{figsigel}
\end{figure}

In Fig.\ref{figsigel} we show the frequency dependence of real and imaginary parts of
(a) $\sigma_{xx}$, $\sigma_{xy}$ and (b) $\varepsilon_{xx}$, $\varepsilon_{xy}$ which are
the only non-vanishing components for the incident radiation linearly polarized along
$x$ axis and propagating perpendicular to the 2DEG. The bulk dimension of
$\sigma_{ij}$ in ${\rm sec^{-1}}$ corresponds to our understanding of the system as having
a finite layer thickness parameter which enters below in the corresponding expressions for
the Faraday and Kerr rotation angles which typically include such a characteristic of the
system as the traveling distance or thickness.
By analyzing Fig.\ref{figsigel} one can see that the components of both
the conductivity and the dielectric tensors reach their local maximum values at frequencies
corresponding to the distance between centers of the magnetic subbands shown
in Fig.\ref{figen}. One can see that the induced off-diagonal component $\varepsilon_{xy}(\omega)$
of the dielectric tensor at certain frequencies can be quite significant and reach
the magnitude of the order of $0.5$ which is a sizable off-diagonal contribution to the static
dielectric constant which is about $12.5$.\cite{Goldberg}
The detailed structure of the conductivity plots in Fig.\ref{figsigel}(a) near $\nu=0.28$ THz where some
 of the conductivity components change their sign is marked by rectangle and will be discussed below.
The practical value of the results shown in Fig.\ref{figsigel} is well-known: since the measurements
 of the radiation-induced conductivity are among the most popular experimental tools for determining
 the parameters of heterostructures, our calculations provide a useful prediction of experimentally measurable
quantities suitable for description of actually used structures with strong SOC.

\begin{figure}
  \centering
  \includegraphics[width=85mm]{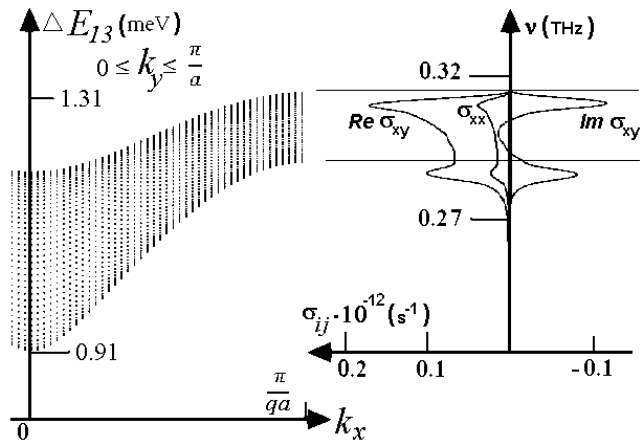}
  \caption{(Left) The $k_x$-dependence of the energy difference $E_{13}(k_x,k_y)=E_3(k_x,k_y)-E_1(k_x,k_y)$
           between the subbands 1 and 3 for to the frequency range on the inset in Fig.\ref{figsigel}
           and (Right) the corresponding photon energy dependence of the conductivity tensor components.
           The points where the ${\rm Im}\sigma_{xy}$ component changes its sign correspond to
           the areas in $E_{13}$ plot where the curvature of the $E_{13}(k_x,k_y)$ function changes its sign.}
  \label{figdifsig}
\end{figure}

The detailed structure of the photon energy dependence of the conductivity and dielectric tensor components
shown in Fig.\ref{figsigel} can be clarified to some extent if we consider it together
with the energy dispersion relations in those subband pairs which are relevant for
the given part of the photon frequency range.
In Fig.\ref{figdifsig} we show the $k_x$-dependence of the energy difference
$E_{13}(k_x,k_y)=E_1(k_x,k_y)-E_3(k_x,k_y)$
between the subbands 1 and 3 corresponding to the frequency range on the inset in
Fig.\ref{figsigel}. The whole energy range is covered by running the $k_y$ component
over half on the Brillouin zone range $0 \le k_y \le \pi/a$, and the $k_x$ component in Fig.\ref{figdifsig}
varies between $0$ and $\pi/qa$ (here $q=1$) since the subband energy spectrum
$E_m(k_x,k_y)$ in the square lattice with Rashba SOC is invariant under
the transformations $k_{x,y} \to -k_{x,y}$.\cite{DP}
By comparing the energy dispersion on the left side of Fig.\ref{figdifsig} with
the conductivity frequency dependence on the right side plotted on the same energy
scale one can connect some specific points of the conductivity plots with the properties of
the energy difference dispersion relation. In particular, the points where the imaginary
part of $\sigma_{xy}$ component responsible for the direction of Kerr angle changes
its sign clearly corresponds to the areas in the plot for the energy difference function $E_{13}=E_3-E_1$
for the subbands $1$ and $3$, respectively, where the second order
derivatives, i.e. the curvature of the $E_{13}(k_x,k_y)$ function changes its sign.
Such connections between the conductivity and the dispersion relations are not
uncommon \cite{K4,DP} since the former is determined in (\ref{sigma}) via the matrix elements
of the velocity operators which, in turn, depend on the shape of the dispersion
relation $E_m(k_x,k_y)$ in the particular $m$-th miniband. The example shown in Fig.\ref{figdifsig}
once again demonstrates the importance of the conductivity tensor calculations since its
experimental measurement is possible and may provide us with certain information about
the topological structure of the energy subbands of such non-trivial system as
the 2D superlattice with Rashba SOC in the magnetic field considered here.

\section{Faraday and Kerr rotation angles}

The knowledge of the radiation-induced conductivity tensor $\sigma_{ij}(\omega)$
can be applied for determination of (depending on the setup) the Kerr or Faraday rotation
angles which are often used for experimental characterization of various
heterostructures.\cite{Oppeneer,Wierman,Crooker,Richard,Antonov,Sato,Hamrle,Reichl,Neudert,Kim,Stroppa,Etz}
Since the most straightforward manifestation of the Faraday or Kerr rotation is the
rotation of the polarization plane on a specific angle, we consider in our manuscript
the linearly polarized incident radiation incoming perpendicular to the 2DEG plane $(xy)$
with the electric field vector parallel to the $x$ axis, ${\bf E}=(E_x,0,0)$.
The interaction of external field with the media may generate the $y$-component of the
electric field which corresponds to the rotation angle $\theta$ described simply as \cite{Volkov}
${\rm tan}(\theta)=E_{y}/E_{x}$. It is clear that although this rotation angle arises due to the application
of the external field, its magnitude is not directly related to the strength of the field components
$E_{x,y}$ due to the expression with a fraction where the common field amplitude is cancelled,
and thus can be measurable even for moderated excitation amplitudes. It is known that the rotation angle
can be defined as the rotation per unit distance traveled by the incident wave without taking into account
a specific thickness of the structure.\cite{Magarill} In this paper we apply the term "2DEG"; however,
one should keep in mind that it is essentially contained in a triangular InGaAs/GaAs quantum well with
finite thickness $d$ which value affects the actual rotation angle as the travel distance.
In the following calculations we use a typical value $d = 100$ nm, and we apply an expression for
the Faraday rotation angle derived for a thin film with thickness $d$ (here $d \ll \lambda$ where $\lambda$
is the incoming wavelength) deposited on a GaAs substrate with the index of refraction
$n_s \approx 3.5$ \cite{Goldberg} which has the following form:\cite{Kim,Volkov,Magarill}

\begin{equation}
\theta_F \approx \frac{{\rm Re}(\sigma_{xy})}{\sigma_{xx}}
\left[1+\frac{1}{Z_{+} \sigma_{xx}} \right]^{-1}
\label{Faraday}
\end{equation}

where $Z_{\pm}=d/[c \cdot (n_s \pm 1)]$.
The Faraday effect involves the transmission of the incident radiation, and one has
the real part of the off-diagonal conductivity present in (\ref{Faraday}).
It is clear that in the limit of a very thin 2DEG layer $d \to 0$ or in frequency intervals where
the longitudinal conductivity $\sigma_{xx}$ drops, the Faraday angle is determined solely
by the off-diagonal part of the conductivity tensor,

\begin{equation}
\theta_F (\sigma_{xx} \to 0)\approx {\rm Re}(\sigma_{xy}) \frac{d}{c} \frac{1}{(n_s+1)},
\label{Faraday2}
\end{equation}

in accordance with what has been shown earlier.\cite{Volkov,Magarill}
Then, under the same approximations one can apply also the relation for Kerr angle $\theta_K$
which reads as \cite{Kim}

\begin{equation}
\theta_K \approx \frac{{\rm Im}(\sigma_{xy})}{\sigma_{xx}^2}
\left(-\frac{2c}{d}\right)
\left[\left(1+\frac{1}{Z_{+} \sigma_{xx}} \right)
      \left(1+\frac{1}{Z_{-} \sigma_{xx}} \right)
\right]^{-1}
\label{Kerr}
\end{equation}

and is determined by the reflection, i.e., imaginary part of the off-diagonal
component of the conductivity tensor.

\begin{figure}
  \centering
  \includegraphics[width=85mm]{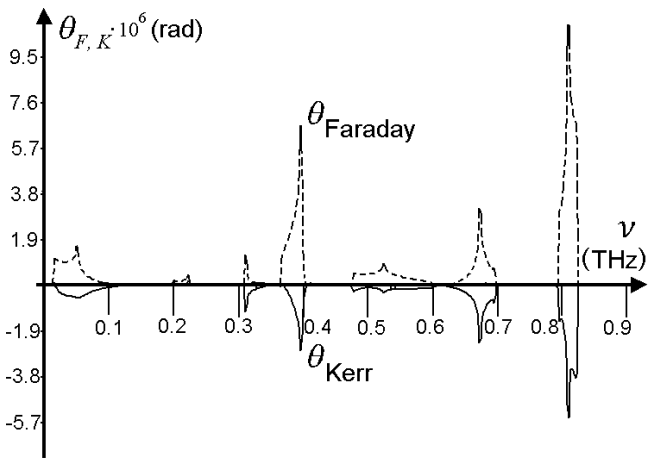}
  \caption{Frequency dependence of the Faraday (dashed curve) and Kerr (solid curve) rotation angles for
           the 2DEG with Rashba SOC described by the conductivity tensor with components shown in Fig.\ref{figsigel}.
           The peaks of both Faraday and Kerr rotation angle correspond to the intervals where at the same frequency
           the longitudinal conductivity $\sigma_{xx}(\omega)$ is small and the off-diagonal part
           $\sigma_{xy}(\omega)$ is finite, i.e. for the samples in the insulator regime where only
           the transverse current is induced by the external radiation.}
  \label{figfake}
\end{figure}

It is clear that here both Faraday and Kerr angles are determined by
the conductivity tensor components only, and thus the frequency dependence of
$\sigma_{ij}(\omega)$ discussed above and shown in Fig.\ref{figsigel}
can be directly applied for the calculation of the frequency dependencies of the Faraday and Kerr angles which
are shown in Fig.\ref{figfake}. A detailed comparison of Fig.\ref{figsigel} and Fig.\ref{figfake}
leads to the conclusion that the peaks of the Faraday or Kerr rotation angle correspond mainly
to the intervals where at the same frequency the longitudinal conductivity $\sigma_{xx}(\omega)$
is small and the off-diagonal part $\sigma_{xy}(\omega)$ is finite, i.e. for
the samples in the insulator regime where only the transverse current is induced by the external
radiation. The opposite sign of the Kerr angle compared to the Faraday angle which can
be observed on the dominating part of the frequency range in Fig.\ref{figfake} can be explained
by drawing the attention to the opposite signs in the expressions (\ref{Faraday}) and (\ref{Kerr})
reflecting the additional phase rotation during the reflection for the Kerr effect compared to
the absorption for the Faraday effect, and the observation that both ${\rm Re}\sigma_{xy}$
and ${\rm Im}\sigma_{xy}$ have mainly constant signs in the frequency regions when
they are both significantly non-zero with relatively small exceptions such as the region
in Fig.\ref{figdifsig} discussed in detail above. It should be noted that the magnitude of both Faraday
and Kerr angles shown Fig.\ref{figfake} is rather small compared, for example, with the ones for
the interband transitions in ferromagnets or metal-doped semiconductors \cite{Kim,Stroppa,Winter}
while being comparable to the Faraday angle for strained GaAs or InGaAs non-magnetic semiconductor
structures,\cite{Kato}, to the Kerr angle in some layered metal structures \cite{Etz} or organic molecular
semiconductors.\cite{Fronk} Nevertheless, it should be stressed that the frequency dependence of Faraday
or Kerr rotation angles provide a quantitative experimental tool for investigation of such fragile
and sophisticated spectrum as the miniband structure of a superlattice with SOC in high magnetic field.

\section{Induced spin textures}

In systems with SOC the response to the external electromagnetic radiation can be seen
not only in charge but also in spin degrees of freedom. The arising non-equilibrium
distributions of local spin density hereafter called the spin textures\cite{K4} are promising
for further applications in spintronics and we believe that they can be probed, as any other
induced magnetic polarization,  by microscopic magnetization detectors \cite{Awschalom,Zutic,Wu2}
as the areas of local magnetization with spatially varying direction, by the Faraday rotation
measurements \cite{Kato} or Kerr microscopy.\cite{Neudert}
Staying in the framework of the linear response theory\cite{Kubo} one can obtain the excited spin density
in the same fashion as the charge conductivity (\ref{sigma}), namely,

\begin{eqnarray}
S_{i}^{j}(x,y,\omega)=\frac {e E_0}{8\pi m \hbar \omega} \sum\limits_{{\bf k},\mu,\nu}
s^{i}_{{\bf k}\mu \nu}(x,y) v_{{\bf k}\nu\mu}^j  f_{{\bf k}\mu}(1-f_{{\bf k}\nu}) \times \\
\times \delta (E_{{\bf k}\nu}-E_{{\bf k}\mu}-\hbar \omega),
\label{spin}
\end{eqnarray}

where $E_0$ is the electric field amplitude in the incident wave, and the function $s^{i}_{\mu \nu}(x,y)$ does not
include the integration over space and thus cannot be referred as a matrix element of the spin operator but instead
can be described as the position-dependent interband spin density function reading as

\begin{equation}
s^{i}_{{\bf k}\mu \nu}(x,y)=\psi^{\dagger}_{{\bf k}\mu}(x,y) {\hat \sigma}_i \psi_{{\bf k}\nu}(x,y).
\label{spinden}
\end{equation}

The quantity  $S_{i}^{j}(x,y,\omega)$ has the meaning of the local spin density created
in the given point $(x,y)$ of the superlattice by the external electromagnetic radiation with frequency $\omega$ with
polarization $j$ which is fixed in our problem as $j=x$, and therefore this index is omitted below. The results for
induced spin texture components corresponding to two peaks of the conductivity (Fig.\ref{figsigel}) or rotation angle
(Fig.\ref{figfake}) frequency dependencies for $\nu = 0.43$ THz and $\nu = 0.64$ THz are shown below in Fig.\ref{figspin}
and Fig.\ref{figspin2}, respectively, in one superlattice cell $-a/2 \le x(y) \le a/2$ normalized on the superlattice
unit cell area $a^2$, i.e. the magnetic moment being actually measured by the probe with area $dS$ can be obtained
after multiplying the spin density by a factor $dS/a^2$. The spin textures are shown in units for the degree of carrier
polarization (i.e. in units of Bohr magneton per carrier) for the incident power of $1.0$ ${\rm mW/cm^2}$ which
is accessible in modern experiments with nanostructures. It can be seen that all components of the induced spin
density are excited on a comparable scale shown in figures (a)-(c), correspondingly which makes these predicted
space distributions promising for the experimental observations.

\begin{figure}
  \centering
  \includegraphics[width=85mm]{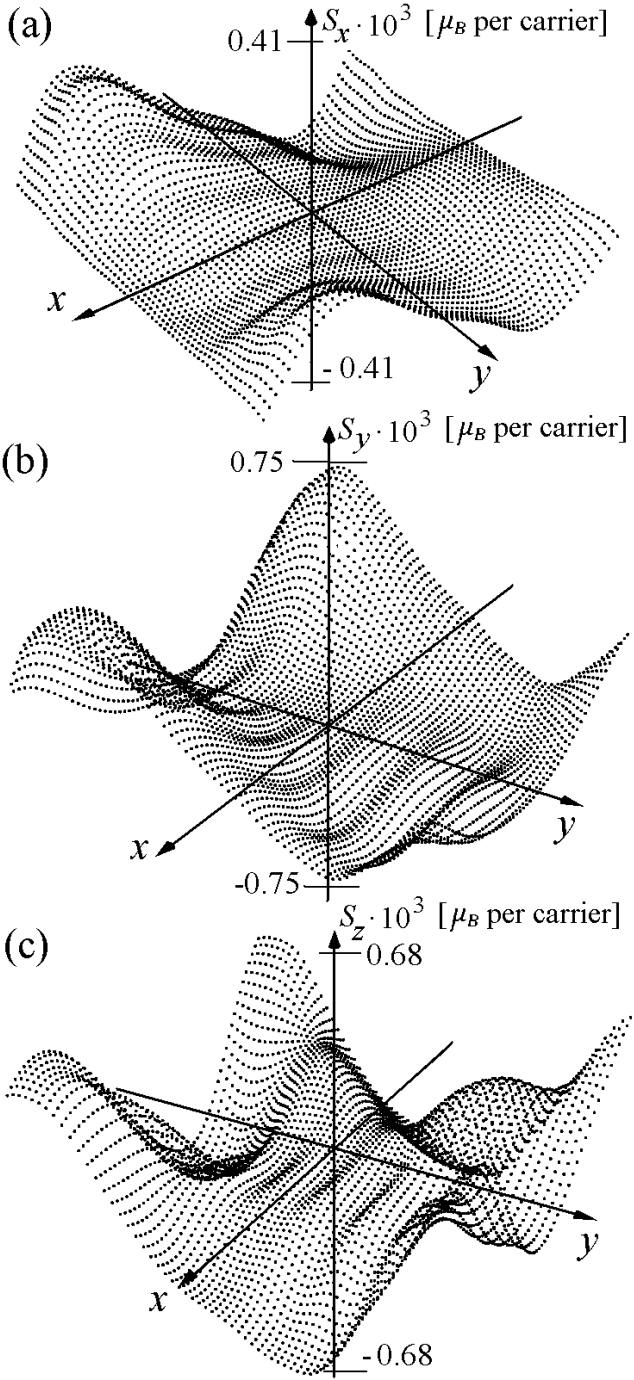}
  \caption{Radiation-induced spin textures shown in units for the degree of carrier polarization
           for the incident power of $1.0$ ${\rm mW/cm^2}$: (a) $S_x(x,y)$, (b) $S_y(x,y)$
           and (c) $S_z(x,y)$ plotted in one superlattice cell $-a/2 \le x(y) \le a/2$ and
           normalized on the superlattice unit cell area. The presence of static magnetic
           field along $z$ direction together with the Rashba SOC in the $(xy)$ plane creates the excited spin textures
           with all spin components being of a comparable order. The dominating space scale of the spin texture
           shape is comparable with the superlattice size which is the lateral size of the figure.}
  \label{figspin}
\end{figure}

The explanation of the rich spin texture structure can be drawn if we take into account the presence of static
magnetic field along $z$ direction together with the Rashba SOC in the $(xy)$ plane which, in combinations with
the linearly polarized radiation along $x$ axis may indeed create the excited spin textures with all spin components
being non-zero. As to the dominating space scale of the spin texture shape which can be observed in
Fig.\ref{figspin},\ref{figspin2}, one can see that it is comparable with the superlattice size which is the lateral size
of the figure being also the scale of the corresponding wavefunctions shape.\cite{DP} The integration of the spin
density components (\ref{spin}) over the superlattice cell indicates that all mean values for spin texture components
in Fig.\ref{figspin},\ref{figspin2} are very close to zero. This result means that the external radiation which
is treated as a perturbation provides no significant gross change in the magnetization over the whole sample but
it indeed can alter the local magnetization in different parts of the structure.

\begin{figure}
  \centering
  \includegraphics[width=85mm]{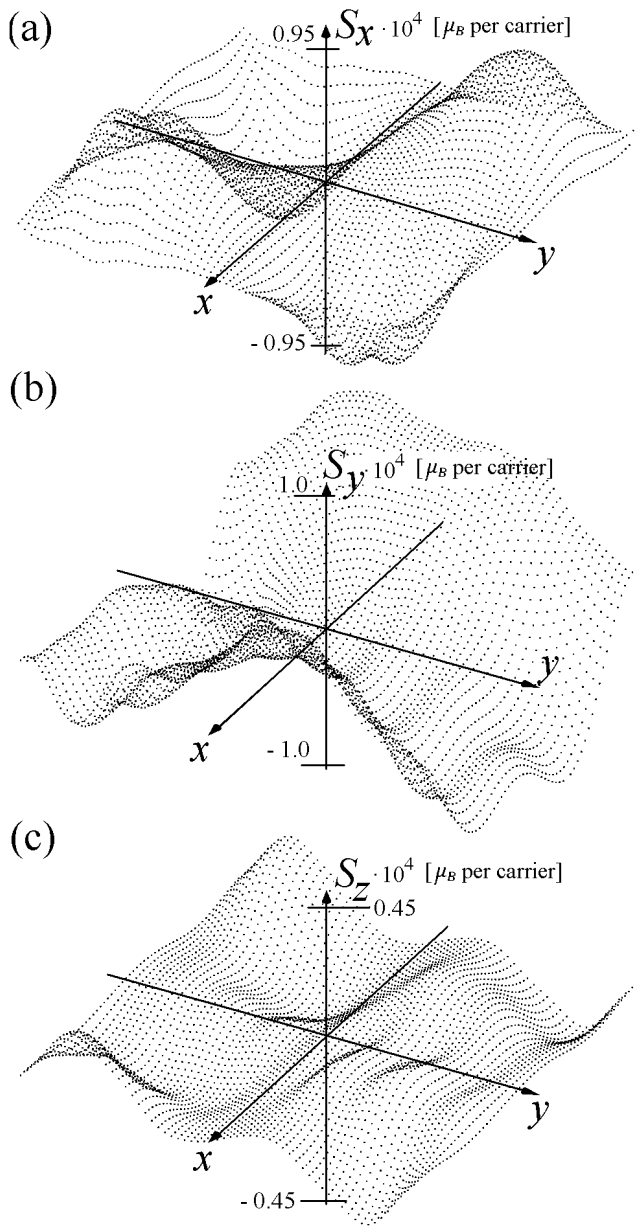}
  \caption{The same as in Fig.\ref{figspin} but for the spin textures excited by the incident radiation with
           the higher frequency $\nu = 0.64$ THz and all other parameters remaining as in Fig.\ref{figspin}.}
  \label{figspin2}
\end{figure}

The transitions corresponding to the higher frequency  $\nu = 0.64$ THz with spin textures in Fig.\ref{figspin2} occur
between the lowest occupied subband 1 in Fig.\ref{figen} and higher subbands compared to the transitions with textures
in Fig.\ref{figspin}, i.e. they are located to the right along the energy axis at the line $p/q=4/1$ in Fig.\ref{figen}.
The calculation of induced spin  textures at different excitation frequencies is also important for possible further
applications of the observed effects at variable parameters. The results for excitation frequency $\nu = 0.64$ THz
are shown in Fig.\ref{figspin2} where again (a) $S_x$ (b) $S_y$ and (c) $S_z$ spin density
components, respectively, are shown on separate plots in one square superlattice cell.

If one compares the induced spin textures in Fig.\ref{figspin} and Fig.\ref{figspin2} for different
excitation frequencies, both common and distinct features can be found. The common
features are the comparable space period and global shape of the excited textures for
all spin components on both figures. This commonalty is stipulated by the common structure
of initial and final quantum states for both examples including the space distribution
of charge and spin density. Besides, for any excitation frequency the resulting excited
spin textures is formed in principal by all allowed transitions between many states below and above the Fermi
level, and thus one may expect certain degree of commonalty between them.

As to the difference between the spin textures at different excitation frequencies in Fig.\ref{figspin}
and Fig.\ref{figspin2}, one can observe the following. First, for the lower frequency
$\nu = 0.43$ THz in Fig.\ref{figspin} all three excited spin components have a higher magnitude.
The explanation can be obtained by taking into account the higher position in energy of the final quantum states
for the induced textures in Fig.\ref{figspin2} which are characterized by more complicated and faster oscillating
wavefunctions,\cite{DP,PS} which, in turn, reduce the magnitude of the corresponding matrix elements and create
the spin density with lower magnitude in Fig.\ref{figspin2} compared to Fig.\ref{figspin}. The same physical mechanism
is responsible for the in general more curved shape of the spin textures in Fig.\ref{figspin2} compared to the ones
in Fig.\ref{figspin}, i.e. the transitions to the quantum states with higher energy leads to the richer space
shape of the excited spin textures.

It should be mentioned again that in Eq.(\ref{spin}) the resulting excited spin textures are formed by all allowed
transitions and by many states below and above the Fermi level and thus have an integral and pretty universal
nature. First of all, this circumstance makes them robust to some extent against the possible corrections
to our model, i.e. the finite sample size, the finite temperatures, the scattering on defects and phonons, and
the presence of other small terms absent in our Hamiltonian. Then, these spin textures can be viewed as measurable
quantities which can be obtained by a probe at a given point $(x,y)$ and thus can be considered as a promising degree
of freedom for further applications such as information processing and storage in spintronics.
It is interesting to note that the features of the excited spin density discussed here were also observed in qualitatively
the same manifestation and with the same space shape in a different system with 2DEG and SOC subjected to periodic
potential of 1D superlattice without the static magnetic field but either under scattering or under radiative
(with various polarizations) or DC electric current excitation \cite{K4} which allows to consider them
as an intrinsic characteristic of low-dimensional systems with strong SOC and non-uniform periodic potential.

\section{Conclusions}

We have studied the terahertz radiation-induced conductivity and dielectric polarization tensors, the Kerr and Faraday
rotation angles, and the excited spin textures for two-dimensional electron gas with strong spin-orbit coupling
subjected to high magnetic field and to tunable periodic potential of a two-dimensional gated superlattice.
It was found that both real and imaginary parts of the frequency-dependent conductivity approach maximum values
with sharp and detectable peaks at frequencies corresponding to the inter-subband transitions between spin-split
magnetic subbands. The observed properties of the conductivity tensor frequency dependence were applied for
the description of the Kerr and Faraday rotation angles which can be used as another experimental tool for describing
the electron gas in periodic structures with significant spin-orbit coupling. The formation of radiation-induced
spin textures is predicted having both in-plane and out-of-plane components with space distribution scale
comparable to the superlattice cell size which can be promising in further experimental and technological applications.

\section*{Acknowledgments}

The authors are grateful to V.Ya. Demikhovskii and A.M. Satanin for helpful discussions.
The work was supported by the RNP Program of Ministry of Education and Science RF
(Grants No. 2.1.1.2686, 2.1.1.3778, 2.2.2.2.4297, 2.1.1.2833), by the RFBR
(Grants No. 09-02-01241a, 09-01-00268a), by the USCRDF (Grant No. BP4M01),
and by the President of RF Grant for young candidates of science No. MK-1652.2009.2.

\end{document}